# Collective intelligence in Massive Online Dialogues

TARANEH KHAZAEI and LU XIAO, University of Western Ontario

1. INTRODUCTION

Collective intelligence refers to the intelligence that emerges from local interactions among individual people. These local interactions in human collectives can be either direct or indirect. In addition, the interactions of both forms can take place through verbal communication or non-verbal action. As illustrated in Figure 1, following these two binary distinctions, interactions among humans can be classified into four groups. Meanwhile, the emergence and ongoing development of Web 2.0 technologies have enabled advanced forms of collective intelligence, allowing large numbers of individuals to act collectively and create high quality intellectual artifacts. To understand the current status in this research and development area, we conducted a systematic review of the studies that used automatic tools to analyze the large-scale Web-based platforms and identified the research gaps. We focused on the platforms in which intelligence may arise from dialogue exchange and discussion among participants (left upper quadrant in Figure 1). Therefore, online reviews, microblogs, and news headlines are excluded since these text units are independent pieces of writings. Chat and email tools are also excluded because they are primarily designed to support small-scale interactions.

To classify the existing literature, the McGrath's framework of group studies [McGrath 1984] is used. Although this framework has been originally intended to address relatively small groups, it is not merely a model about small group activities, but a framework for studying groups systematically. Hence, here, it is applied to research on larger online collectives. In this framework, *group interaction process* is considered the essence of a group and it refers to "patterned relations among the behaviors of individuals" [McGrath 1984]. The framework encompasses several other elements that are *member's properties*, *group structure*, *environment*, and *tasks*. In addition, McGrath describes interaction processes at a micro level, viewing it in terms of three different stages: *communication pattern* among interacting people, *content* of communications, as well as *impact* of different group interaction processes on each other and on participants. The content itself is then viewed in terms of two different aspects: *interpersonal relationship* and *task performance*. Current research on the analysis of collective intelligence commonly treats members' properties, group structure, environment, and task as the context of the study and then facilitates computational tools to analyze the interaction processes. As such, the prior research is categorized into three main groups: communication pattern analysis, content analysis, and impact analysis. Content analysis is then divided into two categories of research. Due to the different nature of virtual spaces, interpersonal relationships component is seen at a higher level, focusing on the studies that analyze sentimental aspects of the content. The task performance component is called task/purpose performance since such settings might be less task-oriented.

1.1 Communication Patterns

Communication pattern refers to the structure of a series of interactive behaviours that takes place among people [McGrath 1984]. A study by Woodley et al. [Woolley et al. 2010] has shown that the patterns of turn taking among participants is correlated with the collective intelligence. This finding illustrates the importance of the analysis of communication patterns in understanding and fostering intelligence. Even though earlier works in this context have relied on basic statistical measures [Whit-taker et al. 1998; Smith 1999], more sophisticated computational techniques have been proposed. A large number of studies first establish a social network of participants and then employ various graph analysis methods proposed by social network analysts [Sack 2000; Gómez et al. 2008; Laniado et al. 2011; Wu et al. 2011; Rangwala and Jamali 2010; Adamic et al. 2008]. Rather than forming social networks from conversations, some researchers have explored communication patterns as information cascades or threads [Gómez et al. 2008; Laniado et al. 2011; Kumar et al. 2010; Gómez et al. 2011; Gómez et al. 2013; Sack 2000]. In these studies, the cascades are normally modeled as a tree, where each node represents a message by a participant, and there is an edge between two nodes if one of these messages is in response to or influenced by another. In addition, some attempts have been made to characterize individuals' communication behaviours such as modeling the time intervals of their activities [Vaz de Melo et al. 2013] or the use of ego networks [Adamic et al. 2008; Welser et al. 2007].

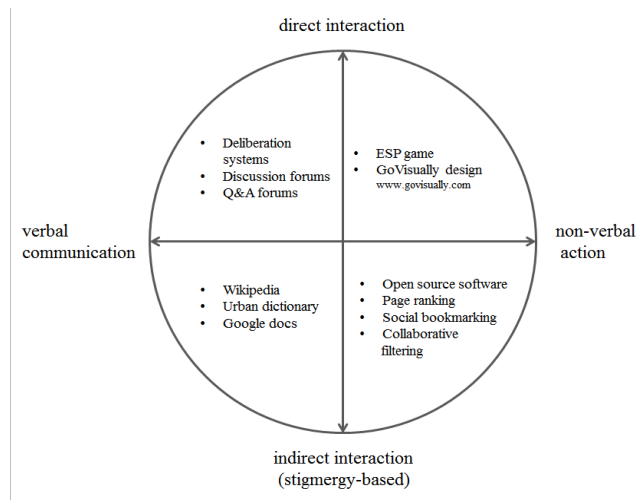

Fig. 1: Collective intelligence can be categorized into four different groups based on the interaction types among humans.

1.2 Content Analysis

1.2.1 *Sentiment.* There has been a large amount of work on sentiment analysis, with researchers investigating various methods to identify subjective sentences, to rate the sentiment level, to detect moods and emotions, and finally, to understand the source, target, and complex attitude types [Pang and Lee 2008]. A traditional lexicon-based approach uses sentiment-annotated dictionaries to determine the sentiment score of a text unit by averaging out sentiment values of individual words extracted from the dictionary [Wanner et al. 2011; Li and Wu 2010]. The increasing availability of labeled data on the Web has led to the use of a variety of supervised classification methods in sentiment research [Rosenthal and McKeown 2012; Sood and Churchill 2010]. Following a related approach, a language model is constructed in [Hassan et al. 2010] to detect sentences with attitudes.

1.2.2 *Task/Purpose Performance.* A set of studies on online environments attempts to address the information overload problem by describing the features of the underlying social space. Such studies can be seen as preliminary steps that can lead to easier and more accurate task performance analysis by machines or users. Adoptions and extensions of document clustering techniques [Said and M. Wanas 2011; Paukkeri and Kotro 2009] as well as topic modeling methods [Zhu et al. 2008] have been proposed to describe content-based features of the conversational text. Another line of work on descriptive analysis has focused on the annotation of the communication space with social actions and behaviours that are of value in performance analysis such as identification of claims [Rosenthal and McKeown 2012; Marin et al. 2011], agreements or disagreements [Abbott et al. 2011; Murakami and Raymond 2010], justifications [Biran and Rambow 2011], and ideas [Convertino et al. ]. Rather than descriptive analysis, some researchers have developed computational methods to directly evaluate the user-generated content. Variations of information retrieval techniques [Sood and Churchill 2010; Feng et al. 2006; Wanas et al. 2009] as well as supervised learning methods [Weimer et al. 2007; Chai et al. 2011; Wanas et al. 2008; Shah and Pomerantz 2010; Harper et al. 2008] have been proposed to evaluate users' contributions.

### 1.3 Impact

Impact analysis aims to understand the potential effects of group interaction processes on each other and on participants. Employing computational methods, two different procedures have been followed to study impact in online social environments. A set of works have focused on calculating variables of interest independently and then they use basic statistical methods to assess how they correlate [Diakopoulos and Naaman 2011; Chmiel et al. 2011a; De Liddo et al. 2011; Adamic et al. 2008; Laniado et al. 2011]. A few recent approaches utilized more complex methods such as clustering to directly assess the potential impact of one variable on another [Chmiel et al. 2011b].

### 1.4 Discussion

Table I provides a summary of the studies reviewed in this article. Exploring the table may enable researchers to identify current research gaps and to gear their research efforts toward addressing current shortcomings. Such gaps include lack of diversity in environments of focus in sentiment analysis, lack of focus on task-oriented environments such as deliberation and idea management tools, as well as lack of sophisticated computational tools to analyze impact. As can be seen in the table, some attempts have been made to analyze the impact of interaction processes, i.e., communication pattern, task/purpose performance, and sentiment, on each other. However, less attention has been paid to understand how these processes influence participants' individual and collective behaviours over time, such as their perception, learning, and judgement. In order to gain insights into how intelligence emerges from within social interactions and to determine various factors that may influence the collective intelligence phenomenon, further research is required to fill these gaps.

Table I. : Prior research is summarized according to the McGrath's framework.

| Primary Purpose | Aspects of Focus | Main Methods | Environments | Secondary Benefits |
|---|---|---|---|---|
| communication analysis | interaction structure<br>thread structure<br>interaction behaviour | social network analysis<br>tree analysis<br>machine learning | comment sets<br>social networks<br>discussion fora<br>Q&A fora | browse & navigation<br>popularity detection & prediction<br>controversy detection & prediction |
| content:<br>sentiment analysis | subjective polarity & rating<br>emotions and moods<br>attitudes | lexicon-based<br>machine learning<br>language modeling | discussion fora<br>comment sets | understanding social relations<br>social action/behaviour detection<br>community management<br>marketing<br>browse & navigation |
| content:<br>performance analysis | topicality<br>social actions & behaviours<br>quality & relevance | clustering & topic modeling<br>supervised machine learning<br>information retrieval | comment sets<br>social networks<br>discussion fora<br>Q&A fora<br>deliberation tools<br>idea management tools | search & navigation<br>marketing<br>technology impact assessment<br>content filtering & summarization |
| impact analysis | performance (topicality) & communication<br>performance (topicality) & sentiment<br>performance (quality) & sentiment<br>sentiment & sentiment<br>communication & learning | basic statistics<br>clustering | comment sets<br>discussion fora<br>Q&A fora<br>learning fora | understanding social<br>& psychological processes<br>technology impact assessment<br>community management |